# Systemic Risk Identification, Modelling, Analysis, and Monitoring: An Integrated Approach

*Antoaneta Sergueiva*[*]

University College London, Department of Computer Science,

Financial Computing and Analytics Group

a.sergueiva@ucl.ac.uk

(working paper, first draft May 2013)

*Abstract:* Research capacity is critical in understanding systemic risk and informing new regulation. Banking regulation has not kept pace with all the complexities of financial innovation. The academic literature on systemic risk is rapidly expanding. The majority of papers analyse a single source or a consolidated source of risk and its effect. A fraction of publications quantify systemic risk measures or formulate penalties for systemically important financial institutions that are of practical regulatory relevance. The challenges facing systemic risk evaluation and regulation still persist, as the definition of systemic risk is somewhat unsettled and that affects attempts to provide solutions. Our understanding of systemic risk is evolving and the awareness of data relevance is rising gradually; this challenge is reflected in the focus of major international research initiatives. There is a consensus that the direct and indirect costs of a systemic crisis are enormous as opposed to preventing it, and that without regulation the externalities will not be prevented; but there is no consensus yet on the extent and detail of regulation, and research expectations are to facilitate the regulatory process. This report outlines an integrated approach for systemic risk evaluation based on multiple types of interbank exposures through innovative modelling approaches as tensorial multilayer networks, suggests how to relate underlying economic data and how to extend the network to cover financial market information. We reason about data requirements and time scale effects, and outline a multi-model hypernetwork of systemic risk knowledge as a scenario analysis and policy support tool. The argument is that logical steps forward would incorporate the range of risk sources and their interrelated effects as contributions towards an overall systemic risk indicator, would perform an integral analysis of sources and their interrelations through rigorous mathematical formulation of models capturing quantitative and qualitative information, and would develop a domain representation framework based on knowledge engineering. The incorporation of a range of risk sources involves identification of relevant data scope and availability, the integral analysis requires the formulation of multi-level models and risk measures, the instantiation of the hypernetwork in the systemic risk domain allows formal representation of the evolving understanding of systemic risk. Thus the outlined project aligns with the focus of current and anticipated international research initiatives and contributes to the effort towards systemic risk evaluation and regulation.

[*] The author is grateful to J. Doyne Farmer and Yaneer Bar-Yam for their constructive comments and time, and to Kevin James for the opportunity to attend and present at the Bank of England seminars on systemic risk and financial stability. Would like to thank Jeffrey Johnson for kindly providing a copy of his forthcoming book in advance, and to Marzena Rostek for sending a recent unpublished version of her working paper.

Research capacity is critical in understanding systemic risk and informing new regulation. As Lo (2011) concludes, banking regulation has not kept pace with all the complexities of financial innovation. The academic literature on systemic risk is rapidly expanding and building capacity. A Google Scholar search for *financial "systemic risk"* results in 2,680 papers for the first half of 2013 and 15,900 publications since 2010. A range of risk sources and mechanisms for cascading failure have been identified and models proposed. The majority of papers analyse a single source or a consolidated source of risk and its effect. A fraction of publications quantify systemic risk measures or formulate penalties for systemically important financial institutions that are of practical regulatory relevance. Systemic risk of a banking system, based on interbank lending, has been communicated as the probability distribution of losses to the system caused by a systemic event and the cascading failures it triggers, suggesting a system fragility quantification as the conditional value at risk of the systemic loss distribution (Martinez-Jaramillo et al., 2010). A combination of network centrality measures, for a single-layer network, has been applied to identify systemically important banks (Martinez-Jaramillo et al., 2012), and eigenvector network centrality has been used to identify systemically important financial institutions in the credit default swaps market (Markose at al., 2012). Network centrality relates to the interconnectedness component in the Basel Committee on Banking Supervision (BCBS) guidelines for assessing systemic importance (BCBS, 2011, 2013). Capital surcharge to be levied on systemically important banks has been suggested, based on eigenvector centrality of single-layer networks, in order to internalize systemic risk costs arising from banks' centrality and to mitigate potential socialized losses (Markose at al., 2012, 2013). The challenges facing systemic risk evaluation and regulation still persist as the definition of systemic risk is somewhat unsettled and that affects attempts to provide solutions (Schwarcz, 2011), the only common factor of various definitions being that a trigger event as an economic shock or institutional failure causes a domino effect of bad economic consequences. Our understanding of systemic risk is evolving, and this is reflected in the focus of major international research initiatives (Farmer et al., 2013), as the EU FP7 project CRISIS Complexity Research Initiative for Systemic InstabilitieS focused on better modelling and understanding of financial system and macroeconomic risk and instability, and the EU FP7 Flagship project FuturICT focused on developing new methods to integrate different scientific models, data and concepts, in order to create a paradigm shift from system components towards evaluating their interactions and emergent collective dynamics. The evolving understanding of systemic risk reveals what data are relevant for evaluations, and informs decisions on creating new regulatory databases maintained by national and international regulatory authorities and collected from financial institutions. Therefore, a further challenge to systemic risk evaluation is that awareness of data relevance is rising gradually, some data relevance may not yet be revealed or relevant data may not yet be available or have only been collected over a short recent period. Thus research should also answer database questions regarding data and data frequencies that will support effective systemic risk evaluation and regulation. Finally, though there is widespread support for using regulatory policy to reduce systemic risk, regulation is still in its initial stages. There is a consensus that the direct and indirect costs of a systemic crisis are enormous as opposed to preventing it, and that without regulation the externalities will not be prevented as market participants are directly motivated to protect themselves but not the financial system as a whole, and the effects can spread to the real economy (Anabtawi and Schwarcz, 2011). However, there is no consensus yet on the extent of regulation or even on the formulation of a systemic risk tax, and research expectations are to facilitate the regulatory process. This report outlines an integrated approach for systemic risk evaluation based on multiple types of interbank exposures through innovative modelling approaches as tensorial multilayer networks, suggests how to relate underlying economic data and how to extend the network to cover financial market information. We reason about data requirements and time scale effects, and outline a multi-model hypernetwork of systemic risk knowledge as a scenario analysis and policy support tool. The argument is that logical steps forward will incorporate the range of risk sources and their interrelated effects as contributions towards an overall systemic risk indicator, will perform an integral analysis of sources and their interrelations through rigorous mathematical formulation of models capturing quantitative and qualitative information, and will develop a domain representation framework based on knowledge engineering, as reasoned in (Serguieva, 2012, 2013). The incorporation of a range of risk sources involves identification of relevant data scope and availability, the integral analysis requires the formulation of multi-level models and risk measures, the instantiation of the hypernetwork in the systemic risk domain allows formal representation of the evolving understanding of systemic risk. Thus the outlined project aligns with the focus of current and anticipated international research initiatives and would contribute to the effort towards systemic risk evaluation and regulation.

    Starting with a focus on the banking system and on networks as a modelling approach, the review of recent articles on networks in systemic risk reveals a trend to gradually

introduce more than one sources of risk into the analysis while no integrated measure has been yet evaluated. Martinez-Jaramillo et al. (2010) propose a simple network model and introduce a banking-system fragility measure. The authors study systemic risk based on consolidated exposure data comprising all interbank deposits, credits and loans. Caccioli et al. (2011) assume abstractly that each bank is characterized by a balance sheet including interbank assets, interbank liabilities, non-interbank deposits, illiquid assets, and the capital buffer. They capture characteristics of a banking system through network features, and investigate the contagion versus stability effects of heterogeneous degree distributions, heterogeneous distribution of assets, and correlations in degree. Markose at al. (2012) analyse interconnected US banks in the market of credit default swaps on residential mortgage backed securities. The authors evaluate a systemic risk penalty based on this source and justify it in a regulatory context as a 'super-spreader' tax. Martinez-Jaramillo et al. (2012) extend their earlier technique to produce a more comprehensive analysis of network characteristics and a new measure for identifying systemically important institutions. They consider as risk sources the interbank lending exposures and the payment system flows, analyse a network for each of the two sources, compare network structure, and construct a centrality measure for each network based on several theoretical centrality measures. The authors emphasize the different roles the same banks play in the two networks, however do not suggest an integrated approach. Langfield et al. (2012) investigate a more granular database, build consolidated exposure networks and consolidated funding networks, and then breakdowns by instrument, and by further criteria like instrument maturity or underlying and bank type. The authors produce statistics by network and recognise the significant difference in network structure for different instruments. As a step towards an integrated analysis, they apply a simple clustering technique and visualise a consolidated inter-cluster exposures network and a consolidated inter-cluster funding network. The clustering is based on factors like unsecured lending, secured lending, marketable securities, net credit default swaps sold, securities financing transactions, and derivatives exposure. As a note, these instruments are also the elements of the consolidated interbank exposures, while interbank funding consolidates unsecured lending, secured lending, and repo. No integrated measure of systemic risk is suggested, neither an integrated network structure detailing links between networks. Markose (2013) re-emphasizes the use of the eigen-pair method for systemic risk analytics in evaluating system stability and in ranking systemically important institutions, and reasons on introducing a progressive Pigou tax called super-spreader tax as a stabilizing mechanism. The author presents a comprehensive review of systemic risk research and recognizes in a stylized form the broader scope of interconnections necessary to consider in order to soundly support a macroprudential policy. The paper directs towards possible use of hypergraphs though no integrated structure and measure are evaluated.

An evident feature of systemic risk research is the dependence on data availability. Identifying relevant data, data collection from financial institutions, and developing databases is an ongoing process for regulatory authorities. The importance of understanding systemic risk for maintaining financial stability is widely recognized, however the progress in developing effective approaches is gradual and does not meet the emergency of the task. Martinez-Jaramillo et al. (2010, 2012) argue that this is due to both the difficulty of the task and the lack of necessary data. The authors are only able to perform a comprehensive analysis, as the Mexican central bank has detailed and daily data on the institutions comprising the Mexican banking system, which allows calculating matrices of interbank exposures since January 2005 onwards. The data intensive investigation in Langfield et al. (2012) is only possible due to the new regulatory UK dataset on interbank exposures broken down by counterparty and instrument. The exposures are collected from banks over six-month periods since the first reporting period of July-December 2011, and the dataset is currently collected and maintained by the Prudential Regulation Authority within the Bank of England. According to the authors, this dataset is the most granular representation of a large interbank market available worldwide. The development of the super-spreader tax in Markose at al. (2012) is facilitated by the data on US banks involved in the credit default swaps (CDS) market, as recorded in the Federal Deposit Insurance Corporation Call Reports. Bilateral exposures or netted bilateral exposures are not available, but rather the data for each financial institution are accessible as its market share in terms of gross notional on the sell side of CDS, gross notional on the buy side, gross negative fair value for which the institution is a guarantor, and gross positive fair value for which it is beneficiary. The authors devise an algorithm that randomly allocates the unobserved bilateral exposures, while meeting as constraints the gross values and the existing concentration of market share. The logical and creative analysis in Caccioli et al. (2011) is restricted by not being based on real data. Academic research on systemic risk is largely restricted to publicly available incomplete data, and therefore the analysis of interbank markets makes assumptions regarding unobserved relationships. A number of papers show the significant difference in the analysis and conclusions or even misleading results, when based on complete data versus assumed unobserved relationships, e.g. Mistrulli (2011) and Cont et al. (2013). Research based at regulatory institutions benefits from access to new regulatory datasets related to systemic risk. These are however recently initiated and continuously refined in terms of definition and scope of relevant data. The UK regulatory dataset on interbank exposures, as an example, is initiated with the Recovery and Resolution Plans (FSA, 2011) and the submission templates and instructions are being refined since the first reporting exercise collected by the Financial Services Authority (FSA) in early 2012 through the current fourth data request being collected in mid-2013 by the Prudential Regulation Authority (PRA), while taking into account the feedback from UK banks and the developments at international regulatory authorities as the Financial Stability Board (FSB). Detailed understanding of how banks are interconnected at global level is a key priority for FSB, as initiated with the consultation paper on financial linkages (FSB, 2011) and as recently reported with the successful implementation of phase one of the data gaps initiative (FSB, 2013), including the start in 2013 of the harmonized collection of improved data on exposures for major systemic banks. An international data hub hosted by the Bank for International Settlements (BIS) will hold the confidential data, and participating national supervisory authorities have formed a Governance Group to oversee the pooling and sharing of information. We support the argument that the soundness and effectiveness of developed research methodologies and produced systemic risk analysis depends on the scope of available data.

This report emphasizes that a logical step forward would incorporate a range of risk sources as contributions towards an overall systemic risk measure, and would perform an integral analysis of sources and their interrelations through rigorously formulated mathematical models. Maintaining the focus on network approaches and the banking system, one can present arguments that networks have been prevailingly implemented as visualisation tools to systemic risk rather than as rigorous analytical and knowledge discovery approaches, that single networks or interconnected networks do not meet the task for integral analysis of multiple risk sources and alternative structures as 'networks of networks' or 'multi-level' networks are required, as reasoned in (Serguieva, 2012, 2013). Recent methodology developments can facilitate the work towards constructing approaches with required capabilities. The mathematical formulation introduced by De Domenico et al. (2013), captures complex networks with multiple subnetworks and layers of connectivity. A tensorial framework is formalised to study multilayer networks and to generalise network descriptors. It has not been applied to systemic risk modelling, however it will allow us to adequately expand the interbank network construction in Langfield et al. (2012) and simultaneously incorporate into multilayers risk sources like unsecured lending, secured lending, marketable securities, net credit default swaps sold, securities financing transactions, and derivatives exposure by type of derivative. The formalisation will further allow us to recalculate single-instrument measures, e.g. the centrality measure suggested in Martinez-Jaramillo et al. (2012) and the super-spreader tax presented in Markose et al. (2012), now based on the complex network or based on any scope within that network. Furthermore, the canonical tensors for each layer in the network can be evaluated based on macroeconomic indicators and financial market indicators, while the adjacency tensors also include interbank exposures. Thus a broader underlying structure will affect the observed structure of interbank exposures. Next, each of the instruments now included in the multi-layer network is traded on different markets, and we can zoom into the layer for an instrument and consider banks as trading agents. We can introduce additional layers under that layer by translating the hypergraph market structure accommodating for coexisting exchanges from (Malamud and Rostek, 2013) into the tensorial formalisation. This will allows us within the same structure to simulate the markets that affect systemic risk network formation. Malamud and Rostek (2013) work at the level of markets only and assume strategic traders though do not analyse with real data. On the other hand, the simulation can be based on samples from the Bloomberg database and from the recently developed regulatory SABRE database maintained initially by FSA and now by PRA and containing information on transaction prices, sizes, time, location and counterparty. This will allow identifying from data different types of strategic traders, and improving the quality of simulation. For example, Benos and Sagade (2012) use samples from the Bloomberg and SABRE databases to analyse market structure. Cogently, we consider market structure as contributing to the overall systemic risk structure and affecting the structure in the other layers of the systemic risk network. The markets in some of the instruments can be augmented to include players and strategies explicitly allowing leverage, as the ones described in (Thurner et al., 2012).

Reasonable arguments exist, as final points here, that no approach has been suggested for capturing the evolution of systemic risk structures through time, and that the development is pending of a domain representation framework suitable for instantiation in the broader systemic risk knowledge domain, as reasoned in (Serguieva 2012, 2013). We support the view regarding the evolution of structure that it itself communicates structure. Time-dependent networks are studied in Braha and Bar-Yam (2010) concluding that static network analysis loses valuable information embedded in dynamic networks, that networks follow a "multiscale" dynamics where structure varies significantly both between time scales and over time, and that a more agile strategy of monitoring and centrality evaluation is necessary. The study is based on single-layer social networks, and it is important to examine now the relevance of these conclusions to multilayer interbank exposures networks. The diffusion equations from (De Domenico et al., 2013; Gomez et al., 2013) will be further examined, as they account for intra-layer and inter-layer diffusion and can study time-scales in complex networks, following the argument that interlayer connections can generate new structural and dynamical correlations between components of a system and affect information diffusion. In order to next address capturing the evolution of structure under incomplete information, we will investigate the suitability of (Minku and Yao, 2012) as a knowledge discovery approach to dealing with structural drifts in the multilayer network. The above approaches can contribute to generalising time-dependent multilayer networks and dynamic processes. Further generalisation is introduced through developing a representation framework for the broader systemic risk knowledge domain, expanding beyond the banking system, encompassing the financial sector and linking to the real economy. As a first step, influence beyond the banking system is introduced within the tensorial multilayer network corresponding to interbank exposures. The canonical basis at each layer, i.e. for each type of exposure, can be identified through principal component analysis based on factors including macroeconomic variables. Thus economic structure and cyclicality will affect systemic risk structure. Furthermore, when zooming into the layer for each exposure instrument and presenting the market for that instrument, the corresponding market structure and leverage effects will influence the systemic risk structure. The generalizations described in this paragraph will allow not only representing but also discovering network structure.

Finally, further macro and micro information within the financial sector, the housing market, and other real economy sectors, is relevant to systemic risk and has been effectively explored through various modelling and analytical approaches that do not lend themselves directly as extensions of the tensorial exposures network introduced above but constitute viable current knowledge on systemic risk. Then, we can refer to techniques in knowledge engineering and construct a systemic risk domain where a model is considered as a unit of knowledge, i.e. models are the language for expressing systemic risk knowledge, and the exposures network is one of the models. A multiple–model domain framework is suggested in Serguieva (2004) where models are positioned along dimensions - imprecision being one of them while the other dimensions can be instantiated as meaningful ontologies for a particular domain. The author introduces the imprecision perspective as benefiting from the generalised theory of uncertainty (Zadeh, 2006), and then considers a model as a generalised constraint on information where a different relational type in a constraint communicates different type of uncertainty, and finally formulates multi-perspective generalised constraints to translate the domain into multiple constraints involving multi-perspective relational types unfolding on domain-specific dimensions. It is further reasoned that different combinations of constraints will be able to solve different user-defined queries to the knowledge base varying in scope and focus, however no structure formalisation is suggested for such groupings of models, and the framework is not instantiated in the systemic risk domain. Relevant formalisation here can be introduced through hypernetworks and hypersimplices (Johnson, 2014), by considering each model as a node, each groupings of models as a hypersimplex answering a particular systemic risk query or a policy query, and the domain as a hypernetwork. As systemic risk knowledge evolves, the hypersimplex of models answering the same query this year may differ from the hypersimplex next year, and we are identifying techniques for domain evolution. The purpose is to introduce a rigorous mathematical formalisation for describing and evolving the systemic-risk hypernetwork domain, while still capturing quantitative and qualitative relations. Only then it will transform as an effective decision support or policy support knowledge system facilitating scenario analysis and policy decisions. Importantly, domain evolution brings about the evolution of the underlying data structure.

The description above of the integrated systemic risk approach in first approximation, is followed next by introducing each stage step by step. Starting with the construction of the tensorial interbank exposures network layer by layer, it will cover the following exposures: total prime lending, total issuer risk, total securities financing transactions exposure, total derivatives exposure at default, and total short-term lending. Within total issuer risk a differentiation will be made between marketable securities and net credit default swaps sold, while total derivatives exposure at default will breakdown by underlying asset class, and the total prime lending will be combined with total short-term lending while the focus is on unsecured and secured lending. Thus the following layers are included:

- *unsecured lending* (by counterparty) – the prime lending includes all drawn lending facilities, whether part of a committed facility or not, for which the counterparty has not pledged collateral, and includes wholesale deposits with a maturity of one year or more and all undrawn and committed unsecured facilities; the short-term money placement includes all unsecured placement with maturity less than one year including excess reserves sold at the inter-bank lending rate, such as Fed Funds, Eurocurrency or equivalent, and includes operating accounts, demand deposit accounts, money market deposit accounts, commercial paper, certificate of deposit, bankers' acceptances, commercial paper sweeps, and certificate of deposit sweeps.

- *secured lending* (by counterparty) - all drawn lending facilities, whether part of a committed facility or not, for which the counterparty has pledged collateral; all undrawn and committed secured facilities.

- *marketable securities* (by issuer or obligator) - total mark-to-market (MtM) value of equity and fixed-income instruments reported as positive (long issuer risk) or negative (short issuer risk), excluding asset-backed securities and covered bonds.

- *net credit default swaps sold* (by reference entity) - CDS bought or sold should include both over-the-counter (OTC) and centrally cleared contracts, as the exposure is to the reference entity of the contract not to the counterparty; the net notional CDS (bought and sold) values here are the net of values reported for notional CDS bought and notional CDS sold, and the net value can be positive (long issuer risk) or negative (short issuer risk), while the interest is in the net CDS sold or the short issuer risk.

- *securities financing transactions* (by counterparty) - all securities lending/borrowing as well as repos and reverse repos; gross notional and net exposure are reported in positive terms, where gross notional is for the mark-to-market amount of the securities or cash lent/borrowed and the net exposure is be the amount of exposure net of collateral.

- *interest-rate derivatives exposure at default* (by counterparty) – includes only bilateral OTC interest-rate derivatives exposures rather than those through central counterparties; net MtM before collateral is the total net by counterparty MtM exposure of all OTC interest-risk derivatives positions where these transactions are only netted within netting sets and a netting set is subject to a legally enforceable bilateral netting arrangement; net MtM after collateral is the sum over netting sets of the net MtM before collateral within a set less collateral received within that set, and net MtM after collateral for the counterparty is not necessarily equal to net MtM before collateral less collateral for the counterparty; collateral posted in excess is the sum of all collateral posted (or unreturned collateral) by the reporting bank to a counterparty in excess of the derivatives payables to the counterparty, as this excess captures the resulting credit exposure the reporting bank has to the counterparty for netting sets where the reporting bank is out-of-the-money; banks report interest-rate derivatives exposure at default as the resulting counterparty credit risk exposure net of collateral evaluated either using the MtM method, the standardised method, or the internal model method.

- *FX derivatives exposure at default* (by counterparty) – as above but for derivatives with currency as underlying.
- *credit derivatives exposure at default* (by counterparty) – as above but for the corresponding underlying.
- *equity derivatives exposure at default* (by counterparty.
- *commodities derivatives exposure at default* (by counterparty).

To construct a layer in the network, we follow the tensorial network description in (De Domenico et al., 2013) for a layer $\tilde{k}$:

$$W_\beta^\alpha(\tilde{k}) = \sum_{i,j=1}^{N} w_{ij}(\tilde{k}) E_\beta^\alpha(ij,\tilde{k}); \quad W_\beta^\alpha(\tilde{k}), E_\beta^\alpha(ij) \in \Re^{N \times N}; \tilde{k} = 1,\cdots,L; i,j = 1,\cdots N \qquad (1)$$

where $N$ is the number of nodes $n_i$ in layer $\tilde{k}$ and $L$ is the number of layers. Then, in the context of a national banking system, the number of nodes in each layer is equal to the number of banks in the system, and the number of layers is $L=10$ corresponding to the types of interbank exposures described above. The intensity of the relationship from bank $n_i$ to bank $n_j$ in layer $\tilde{k}$ is indicated with $w_{ij}(\tilde{k})$ and can be evaluated from the exposures along instrument $\tilde{k}$. Next, $E_\beta^\alpha(ij,\tilde{k})$ denotes the $\binom{\alpha}{\beta}$ component of the tensor in the canonical basis for layer $\tilde{k}$, and can be evaluated by identifying macroeconomic indicators and financial

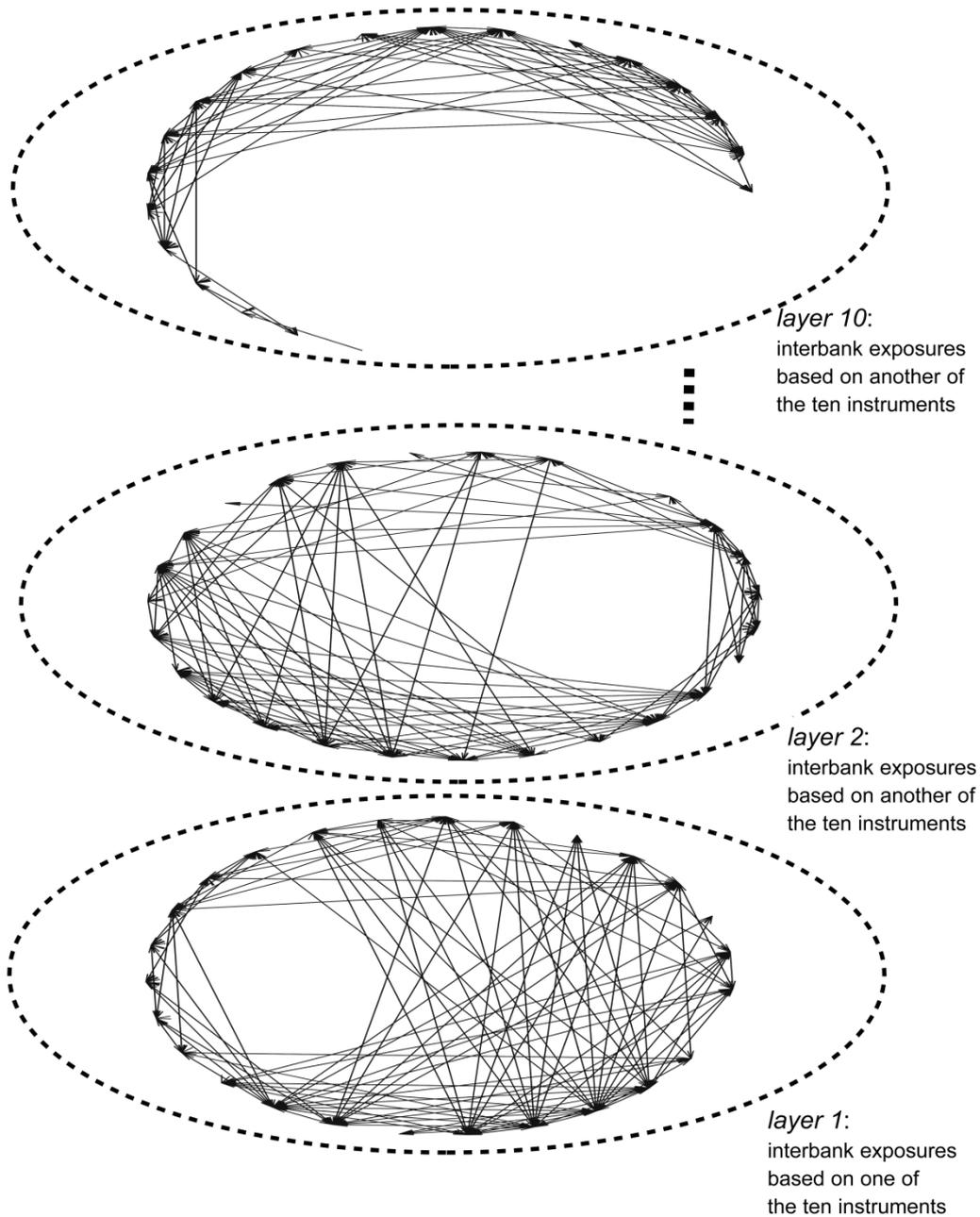

**Fig 1:** *Multiple layers in the interbank exposures network*
Here we first construct the network layer by layer, and then build the interlayer links and link dynamics market indicators correlated with the instrument in layer $\tilde{k}$, then perform principal component analysis of the identified factors and regress this type of exposure along the orthogonal principal components. For example, factors affecting the losses of Mexican financial institutions (Solorzano-Margain et al, 2013), and therefore their exposures, are identified in another context as:

- Dow Jones Industrial Average (DJI)
- Deutscher Aktienindex (DAX)
- Nikkei Heikin Kabuka (Nikkei)
- Bovespa: Brazilian stock market index
- Treasury bill (T-bill) yield with maturities of $t \in \{90, 180, 720, 3600\}$
- London InterBank Offered Rate (LIBOR) yield with maturities of $t \in \{7, 30, 180, 360\}$
- Mexican treasury bond yields without taxes with maturities of $t \in \{28, 91, 182, 364, 720, 1800, 3600, 5460\}$
- Mexican treasury bond yields with taxes with maturities of $t \in \{1, 28, 91, 182, 364, 480, 720, 1800, 3600, 10920\}$
- FX peso-dollar
- Real rate of Mexican bonds against inflation with taxes with maturities of 1,800 days
- Real rate of Mexican bonds against inflation without taxes with maturities of 1,800 days
- Spread on saving protection bonds with maturities of 1,080 days
- Spread on saving protection bonds with quarterly payments with maturities of 1,080 days
- Spread on monetary regulation bonds with maturities of 365 days
- Spread on development government bonds
- Swap internal interbank equilibrium rate yield with maturities of $t \in \{1, 28, 91, 364, 1800, 3600, 7200\}$
- National index of producer prices of non-residential rental buildings
- IPC: index of the Mexican stock exchange

We target to evaluate each stage of the approach for different countries – UK, Italy, Mexico, US – and the above set will be identified for each country. Within a national network, though starting at each layer with the same set of factors, different subsets of them may be correlated to the instrument of that layer and different sets of principal components may be identified at each layer. Thus the tensorial network allows for a richer structure within a layer. Next, $W_{\beta}^{\alpha}(\tilde{k})$ is the $\binom{\alpha}{\beta}$ component of the adjacency tensor for layer $\tilde{k}$. Furthermore, the multilayer adjacency tensor $M_{\beta\tilde{\delta}}^{\alpha\tilde{\gamma}}$ is expressed as:

$$M_{\beta\tilde{\delta}}^{\alpha\tilde{\gamma}} = \sum_{\tilde{h},\tilde{k}=1}^{L} \sum_{i,j=1}^{N} w_{ij}(\tilde{h}\tilde{k})\mathcal{E}_{\beta\tilde{\delta}}^{\alpha\tilde{\gamma}}(ij\tilde{h}\tilde{k}); \quad M_{\beta\tilde{\delta}}^{\alpha\tilde{\gamma}}, \mathcal{E}_{\beta\tilde{\delta}}^{\alpha\tilde{\gamma}}(ij\tilde{h}\tilde{k}) \in \Re^{N\times N\times L\times L};$$
$$\tilde{h},\tilde{k}=1,\cdots,L; \quad i,j=1,\cdots N. \qquad (2)$$

where $\mathcal{E}_{\beta\tilde{\delta}}^{\alpha\tilde{\gamma}}(ij\tilde{h}\tilde{k})$ is the tensors of the canonical basis in the space $\Re^{N\times N\times L\times L}$, $n_i$ is a bank in layer $\tilde{h}$, and $n_j$ is a node in layer $\tilde{k}$. We consider similar approach when evaluating canonical tensors between layers as the one described above about within layers, but will now include financial statement information as factors, as the type of multilayer network will be prevailingly (though not restricted to) multiplex.

The exposure data are available in different countries at different frequencies. In the UK exposure data are collected at 6-month intervals, and in Mexico the availability of daily data is claimed. On the other hand, Braha and Bar-Yam (2010) conclude by evaluating social networks that different nodes have high centrality at different time scales on average. Network centrality measures contribute to identifying systemically important banks, and current regulatory expectations do not exceed evaluating centrality in a static network topology over relatively long periods. We will investigate, subject to data availability, the time scale effects, across countries, across types of exposures and along integrated exposures, and suggest a more agile strategy of monitoring or measure of centrality if the analysis directs towards that. Centrality measures (De Domenico et al., 2013) will be evaluated:

- *degree centrality and strength centrality*

    monoplex: *degree centrality* $d(i)$ *of node* $n_i$ for unweighted $W_{\beta}^{\alpha}$

    *strenght centrality* $s(i)$ *of node* $n_i$ for weighted $W_{\beta}^{\alpha}$

    $$d(i), s(i) = W_{\beta}^{\alpha} u_{\alpha} e^{\beta}(i), \qquad (3)$$

    where $u_{\alpha}$ is 1-vector, $e^{\beta}(i)$ is a contravariant canonical vector

    multilayer: *multi-degree centrality* $D^{\alpha}$

    $$D^{\alpha} = \sum_{h,k=1}^{L} d^{\alpha}(\tilde{h}\tilde{k}) = M_{\beta\tilde{\delta}}^{\alpha\tilde{\gamma}} U_{\tilde{\gamma}}^{\tilde{\delta}} u^{\beta}, \qquad (4)$$

    where $d^{\alpha}(\tilde{h}\tilde{k})$ is the interlayer degree centrality vector corresponding to layers $\tilde{h}$ and $\tilde{k}$, and $U_{\tilde{\gamma}}^{\tilde{\delta}}$ is 1-tensor.

    multilayer: the weighted adjacency tensor $\Psi_{\tilde{\delta}}^{\tilde{\lambda}} \in \Re^{L\times L}$ can be used as a

basis to formulate a strength measure, where

$$\Psi_{\tilde{\delta}}^{\tilde{\gamma}} = M_{\beta\tilde{\delta}}^{\alpha\tilde{\gamma}} U_{\alpha}^{\beta} = \left[ \sum_{\tilde{h},\tilde{k}=1}^{L} \sum_{i,j=1}^{N} w_{ij}(\tilde{h}\tilde{k}) \mathcal{E}_{\beta\tilde{\delta}}^{\alpha\tilde{\gamma}}(ij\tilde{h}\tilde{k}) \right] U_{\alpha}^{\beta} = \sum_{h,k=1}^{L} q_{\tilde{h}\tilde{k}} E_{\tilde{\delta}}^{\tilde{\gamma}}(\tilde{h}\tilde{k})$$

and $q_{\tilde{h}\tilde{k}}$ sums weights in connections between layers $\tilde{h}$ and $\tilde{k}$.

projected monoplex: degree and strength centrality can be formulated for the projected monoplex obtained from the multilayer network, where the projected network is described with $P_{\beta}^{\alpha} = M_{\beta\tilde{\delta}}^{\alpha\tilde{\gamma}} U_{\tilde{\gamma}}^{\tilde{\delta}}$.

- *eigencentrality*

    monoplex: *eigencentrality vector* 
    $$v_{\beta} = \lambda_1^{-1} W_{\beta}^{\alpha} v_{\alpha} \qquad (5)$$

    where $\lambda_1$ is the largest eigenvalue of $W_{\beta}^{\alpha}$ and $v_{\alpha}$ is its corresponding eigenvector.

    multilayer: *eigecentrality tensor* 
    $$V_{\beta\tilde{\delta}} = \lambda_1^{-1} M_{\beta\tilde{\delta}}^{\alpha\tilde{\gamma}} V_{\alpha\tilde{\gamma}} \qquad (6)$$

    where $\lambda_1$ is the largest eigenvalue of $M$ and $V_{\alpha\tilde{\gamma}}$ is the corresponding eigentensor.

Further network centrality measures as *betweenness*, *closeness*, and *Page Rank centrality* (Martinez-Jaramillo, 2012) will be translated for tensorial networks – monoplex, projected and multilayer - and evaluated across countries and time scales. A single centrality measure will be formulated based on the different types of centrality, and its robustness will be evaluated along time scales. A more agile strategy of monitoring will be proposed, if the analysis reveals necessary, and a good indicator will be proposed of a bank's systemic importance.

Existing proposed taxes or capital surcharges for banks spreading systemic risk can be translated for the tensorial network, and new will be proposed now based on the wider approach. For example the tax used in (Markose et al., 2012), based on one instruments as a cause of systemic risk, and aiming at banks with high eigencentrality to internalize the costs their failure will cause to others and to mitigate their impact on system instability. When exposures in a layer of the tensorial network are calculated as proportion of total capital of exposed banks, and system instability is given by the largest eigenvalue $\lambda_1$, and networks are directed along exposures, then we have to use the Katz centrality modification of eigencentrality for monoplex where $\lambda_1 > a$:

$$v_\beta = \left(\Delta_\beta^\alpha - aW_\beta^\alpha\right)^{-1} u_\alpha$$

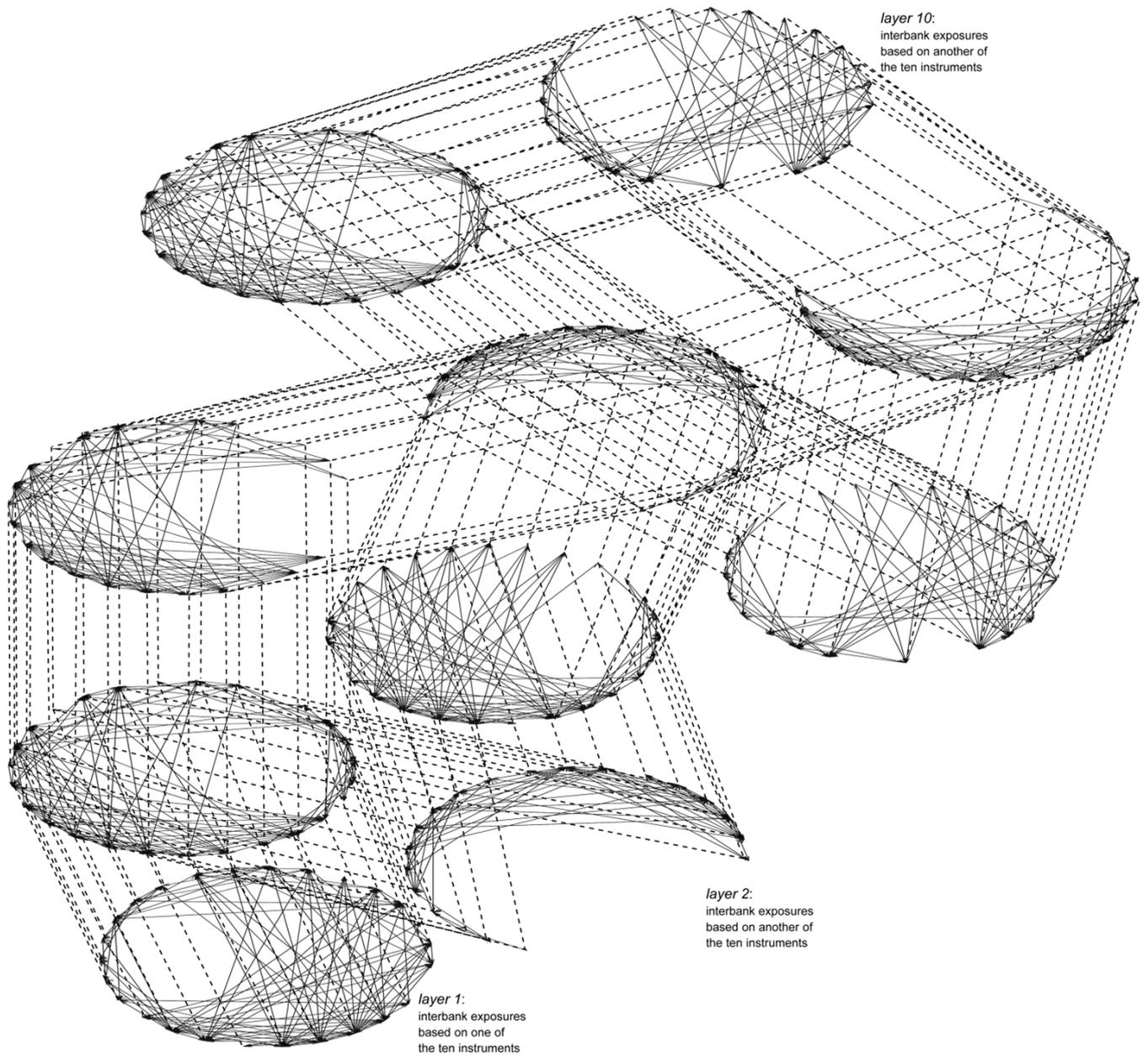

**Fig 2:** *Multiplex exposures network* – the structure here will be further extended by unfolding each of these layers into its market.

The capital surcharge for bank *i* is introduced as proportionate to the $i^{th}$ element of the eigenvector $cv_\beta(i)$ and included in the total capital of bank *i*, and the network is re-evaluated to determine the level of c that stabilizes the system. Such levels can be determined for each instrument presented as a monoplex network, and an analogous procedure will be formulated for the projected and the multi-layer network. The next step is to extend the network by zooming into each layer and model the market for the instrument of that layer, while preserving the multilayer tensorial framework, e.g. by incorporating multilayer markets (Malamud and Rostek, 2013) and allowing for leverage strategies (Thurner at al., 2012).

As argued in (Anabtawi and Schwarcz, 2011; Schwarcz, 2011), either banking system structure or financial markets structure can lead to cascading failures, and systemic risk analysis and regulation should focus on both. We are currently considering suitable multiagent, complex network, and ensemble approaches to modelling financial markets, particularly with the purpose to identify structure at different granularity, to discover dynamic structure, and to differentiate between signals in structure change – which ones to react to and avoide a crisis and which ones to note but not over-react to, as reasoned in (Serguieva, 2013). Details will be provided in the next extended version of this report. We expect to be able to contribute to the CRISIS project in this aspect. Finally, models developed throughout this work, and further models of the banking system, the broader financial system and financial markets, the housing market and economic sectors, as relevant to systemic risk, will be carefully selected and (meta)linked within a knowledge hypernetwork on systemic risk serving as a knowledge base for enquiries and policy scenarios. This will align with further new EU research initiatives on developing an economic and financial exploratory through a variety of modelling approaches and investigating all aspects of risk and stability.